# Damage-free single-mode transmission of deep-UV light in hollow-core PCF

F. Gebert[1], M. H. Frosz[2], T. Weiss[2,3], Y. Wan[1],
A. Ermolov[2], N. Y. Joly[2], P. O. Schmidt[1,4,*] and P. St.J. Russell[2]

[1]*QUEST Institute, Physikalisch-Technische Bundesanstalt, 38116 Braunschweig, Germany*
[2]*Max Planck Institute for the Science of Light, Günther-Scharowsky-Str. 1, 91058 Erlangen, Germany*
[3]*4th Physics Institute and Research Center SCoPE, University of Stuttgart, Pfaffenwaldring 57, 70550 Stuttgart, Germany*
[3]*Institut für Quantenoptik, Leibniz Universität Hannover, 30167 Hannover, Germany*
*piet.schmidt@quantummetrology.de*

**Abstract:** Transmission of UV light with high beam quality and pointing stability is desirable for many experiments in atomic, molecular and optical physics. In particular, laser cooling and coherent manipulation of trapped ions with transitions in the UV require stable, single-mode light delivery. Transmitting even ~2 mW CW light at 280 nm through silica solid-core fibers has previously been found to cause transmission degradation after just a few hours due to optical damage. We show that photonic crystal fiber of the kagomé type can be used for effectively single-mode transmission with acceptable loss and bending sensitivity. No transmission degradation was observed even after >100 hours of operation with 15 mW CW input power. In addition it is shown that implementation of the fiber in a trapped ion experiment significantly increases the coherence times of the internal state transfer due to an increase in beam pointing stability.

## References and Links

1. C. W. Chou, D. B. Hume, J. C. J. Koelemeij, D. J. Wineland, and T. Rosenband, "Frequency Comparison of Two High-Accuracy $Al^+$ Optical Clocks," Phys. Rev. Lett. **104**, 070802 (2010).
2. B. Hemmerling, F. Gebert, Y. Wan, D. Nigg, I. V. Sherstov, and P. O. Schmidt, "A single laser system for ground-state cooling of $^{25}Mg^+$," Appl. Phys. B **104**, 583-590 (2011).
3. T. Rosenband, D. B. Hume, P. O. Schmidt, C. W. Chou, A. Brusch, L. Lorini, W. H. Oskay, R. E. Drullinger, T. M. Fortier, J. E. Stalnaker, S. A. Diddams, W. C. Swann, N. R. Newbury, W. M. Itano, D. J. Wineland, and J. C. Bergquist, "Frequency ratio of $Al^+$ and $Hg^+$ single-ion optical clocks; Metrology at the 17$^{th}$ decimal place," Science **319**, 1808-1812 (2008).
4. Y. H. Wang, R. Dumke, T. Liu, A. Stejskal, Y. N. Zhao, J. Zhang, Z. H. Lu, L. J. Wang, T. Becker, and H. Walther, "Absolute frequency measurement and high resolution spectroscopy of $^{115}In^+$ $5s^{21}$ $S_0$-$5s5p$ $^3P_0$ narrowline transition," Opt. Commun. **273**, 526-531 (2007).
5. A. C. Wilson, C. Ospelkaus, A. P. VanDevender, J. A. Mlynek, K. R. Brown, D. Leibfried, and D. J. Wineland, "A 750-mW, continuous-wave, solid-state laser source at 313 nm for cooling and manipulating trapped $^9Be^+$ ions," Appl. Phys. B **105**, 741-748 (2011).
6. Y. Wan, F. Gebert, J. B. Wübbena, N. Scharnhorst, S. Amairi, I. D. Leroux, B. Hemmerling, N. Lörch, K. Hammerer, and P. O. Schmidt, "Precision spectroscopy by photon-recoil signal amplification," Nat. Commun. **5**, 3096 (2014).
7. J. P. Gaebler, A. M. Meier, T. R. Tan, R. Bowler, Y. Lin, D. Hanneke, J. D. Jost, J. P. Home, E. Knill, D. Leibfried, and D. J. Wineland, "Randomized Benchmarking of Multiqubit Gates," Phys. Rev. Lett. **108**, 260503 (2012).
8. H. Häffner, W. Hansel, C. F. Roos, J. Benhelm, D. Chek-al-kar, M. Chwalla, T. Korber, U. D. Rapol, M. Riebe, P. O. Schmidt, C. Becher, O. Guhne, W. Dur, and R. Blatt, "Scalable multiparticle entanglement of trapped ions," Nature **438**, 643-646 (2005).
9. A. Friedenauer, H. Schmitz, J. T. Glueckert, D. Porras, and T. Schaetz, "Simulating a quantum magnet with trapped ions," Nature Physics **4**, 757-761 (2008).
10. J. W. Britton, B. C. Sawyer, A. C. Keith, C. C. J. Wang, J. K. Freericks, H. Uys, M. J. Biercuk, and J. J. Bollinger, "Engineered two-dimensional Ising interactions in a trapped-ion quantum simulator with hundreds of spins," Nature **484**, 489-492 (2012).
11. N. Yamamoto, L. Tao, and A. P. Yalin, "Single-mode delivery of 250 nm light using a large mode area photonic crystal fiber," Opt. Express **17**, 16933-16940 (2009).
12. J. Nold, P. Hoelzer, N. Y. Joly, G. K. L. Wong, A. Nazarkin, A. Podlipensky, M. Scharrer, and P. St.J. Russell, "Pressure-controlled phase matching to third harmonic in Ar-filled hollow-core photonic crystal fiber," Opt. Lett. **35**, 2922-2924 (2010).
13. J. C. Travers, W. K. Chang, J. Nold, N. Y. Joly, and P. St.J. Russell, "Ultrafast nonlinear optics in gas-filled hollow-core photonic crystal fibers [Invited]," J. Opt. Soc. Am. B **28**, A11-A26 (2011).

14. S. Février, F. Gérôme, A. Labruyère, B. Beaudou, G. Humbert, and J. L. Auguste, "Ultraviolet guiding hollow-core photonic crystal fiber," Opt. Lett. **34**, 2888-2890 (2009).
15. J. C. Knight, T. A. Birks, P. St.J. Russell, and D. M. Atkin, "All-silica single-mode optical fiber with photonic crystal cladding," Opt. Lett. **21**, 1547-1549 (1996).
16. F. Benabid and P. J. Roberts, "Linear and nonlinear optical properties of hollow core photonic crystal fiber," J. Mod. Opt. **58**, 87-124 (2011).
17. J. L. Archambault, R. J. Black, S. Lacroix, and J. Bures, "Loss Calculations for Antiresonant Wave-Guides," J. Lightwave Technol. **11**, 416-423 (1993).
18. J. Pomplun, L. Zschiedrich, R. Klose, F. Schmidt, and S. Burger, "Finite element simulation of radiation losses in photonic crystal fibers," Phys. Status Solidi A **204**, 3822-3837 (2007).
19. I. H. Malitson, "Interspecimen Comparison of Refractive Index of Fused Silica," J. Opt. Soc. Am. **55**, 1205-1209 (1965).
20. S. Schneider and G. J. Milburn, "Decoherence in ion traps due to laser intensity and phase fluctuations," Phys. Rev. A **57**, 3748-3752 (1998).
21. R. Ozeri, W. M. Itano, R. B. Blakestad, J. Britton, J. Chiaverini, J. D. Jost, C. Langer, D. Leibfried, R. Reichle, S. Seidelin, J. H. Wesenberg, and D. J. Wineland, "Errors in trapped-ion quantum gates due to spontaneous photon scattering," Phys. Rev. A **75**(2007).
22. Y. Wan, F. Gebert, F. Wolf, and P. O. Schmidt, (in preparation).
23. Y. Colombe, (personal communication, 2013).

## 1. Introduction

Standard solid-core silica optical fibers are ideal for low-loss delivery of single-transverse-mode beams from the visible to the infrared spectral range. There are, however, a number of applications in which single-mode delivery of ultraviolet (UV) light by fiber would be highly desirable. For example, experiments on coherent manipulation of trapped ions for precision spectroscopy and optical clocks [1-6], quantum information processing [7, 8] and trapped ion simulators [9, 10] all require good beam quality and pointing stability, which is normally precisely what single-mode optical fibers can provide. For these applications a loss of a few dB/m is acceptable so long as the transmission remains single-mode and stable over time. A number of problems arise, however, when using standard optical fibers in the UV. Although single-mode guidance can be maintained (for the same core-cladding index step) simply by reducing the core diameter by the ratio of the wavelengths, or by using an endlessly single-mode solid-core photonic crystal fiber (PCF), most glasses become highly absorbing in the UV, and furthermore the transmission degrades over time due to UV-induced color center formation and optical damage in the core. For example, Yamamoto et al. found that in a PCF with a solid silica core the transmission dropped by more than 90% after ~4 hours when using 3 mW CW light at 250 nm [11].

Recent experiments on nonlinear spectral broadening in gas-filled hollow core kagomé-style PCF have shown that these fibers are able to guide ultrashort pulses of UV light with losses of order 3 dB/m [12] and single-mode beam quality at average powers of ~50 μW [13]. Finite element simulations indicate that the light-in-glass fraction in kagomé-PCF is typically <0.01%, which circumvents the problem of UV-induced long-term damage in the glass. Kagomé-PCF can also be made effectively single-mode by decreasing the core size until higher-order modes have significantly higher propagation losses than the fundamental mode. A kagomé-PCF with 2 dB/m loss at 355 nm was recently demonstrated, but due to the relatively large core (~30 μm) it was highly multimode [14].

In this letter we report the fabrication of a series of kagomé-PCFs with core diameters of ~20 μm. The transmission loss was measured at 280 nm using the cut-back technique, and the output beam quality evaluated under varying in-coupling conditions. Simulations of the loss, together with transmission measurements, indicate that it is polarization-dependent and limited by fabrication-induced variations in the thickness of the core-wall surround. In particular, we show for the first time that few-m lengths of these fibers are capable of transmitting continuous wave UV powers of several mW without degradation. Finally, we show that use of kagomé-PCF in a trapped ion experiment significantly increases the coherence times of the internal state transfer due to a reduction in beam-pointing instabilities.

## 2. Theory and methods

The kagomé-PCFs were fabricated using the stack-and-draw technique [15]. Scanning electron micrographs (SEM) of one fiber are shown in Fig. 1. By careful control of the pressure applied to the core during drawing, samples with different core diameters and wall thicknesses could be obtained. For the UV wavelengths considered here (~280 nm), higher order modes typically experience significantly higher losses, resulting in kagomé-fibers that are effectively single-mode.

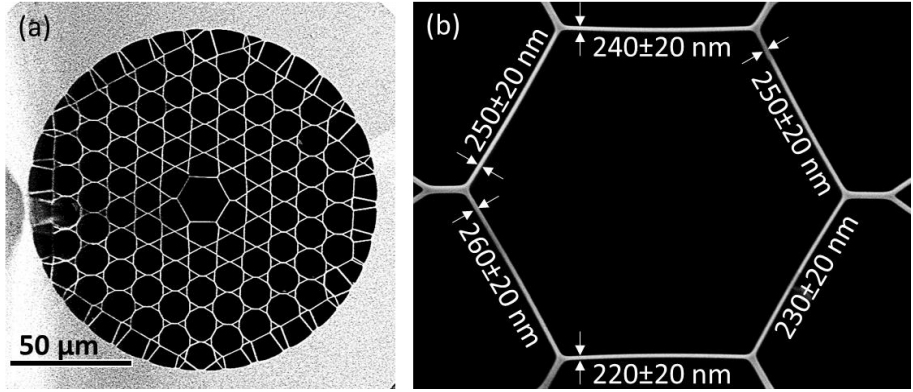

Fig. 1. (a) Scanning electron micrograph (SEM) of fiber sample A. (b) Close-up of the core structure with core-wall thickness measurements.

Kagomé-PCFs guide light with a few dB/m loss over broad transmission windows (several hundred nm), interspersed with narrow bands of high loss [16]. The guiding mechanism in the low-loss regions is not yet fully elucidated, but appears to be a two-dimensional generalization of the ARROW mechanism previously studied in planar and cylindrical structures [17]. The principal high loss bands occur at wavelengths where the core mode phase-matches to modes guided in the core wall, i.e., when [16, 17]:

$$kh_{cw}\sqrt{n_g^2(\lambda) - n_m^2} = m\pi, \quad m = 1,2,3...  \quad (1)$$

where $n_m$ is the modal index (slightly less than 1 for the fundamental mode), $h_{cw}$ is the wall thickness, $k = 2\pi/\lambda$ is the vacuum wavevector and $n_g(\lambda)$ is the wavelength-dependent refractive index of the glass.

We calculated the fiber loss using finite-element modelling (FEM) [18], including the dispersion of the silica glass [19]. To reduce computational complexity the kagomé structure was simplified to only one ring of air-holes around the core (inset of Fig. 2). Comparisons with simulations for the full kagomé-structure showed that this approximation works well, at least within the low-loss windows. It also accurately predicts the position and width of the high loss bands, although not the peak loss values.

At a wavelength of 280 nm Eq. (1) predicts for $m = 2$ a high loss peak at a wall thickness of $h_{cw} = 252$ nm, in agreement with the results of FEM (Fig. 2). For the same parameters, $d\lambda/dh_{cw} \sim 1.1$, indicating that a 1 nm variation in core-wall thickness will cause a ~1 nm shift in the resonance. Since the variations in the actual fibers are much greater, this will cause strong inhomogeneous broadening of the loss peak. Azimuthal variations in core-wall thickness will also break the six-fold symmetry of the structure and cause birefringence and polarization-dependent loss.

As shown in Fig. 1(b), careful analysis of the SEMs shows that the core wall thickness $h_{cw}$ for fiber A varies over the range 240±20 nm and for fiber B over the range 205±15 nm. A fixed wavelength of 280 nm and a core diameter of $2r_{co} = 18.7$ μm (corresponding to fiber A) was used in the FEM, which were for an ideal structure with a constant core-wall thickness. A realistic structure with azimuthal variations in core-wall thickness may display additional losses. The results are shown in Fig. 2. Within the measured range of core-wall thicknesses for fiber A, the loss can potentially reach above 100 dB/m. For fiber B, the calculated maximum loss can reach slightly above 1 dB/m.

In Fig. 2 the main loss peak centered at $h_{cw} \sim 255$ nm corresponds to the $m = 2$ solution of Eq. (1). The additional loss peaks are caused by phase-matching to resonances in the complex cladding structure beyond the core surround and cannot be modeled using the simple approximations used in deriving Eq. (1). The closely spaced loss peaks found in Fig. 2 demonstrate that even though certain values of core-wall thickness would theoretically give very low loss (<0.001 dB/m), the loss will in practice be higher due to nm-scale variations in core-wall thickness in the actual fibers.

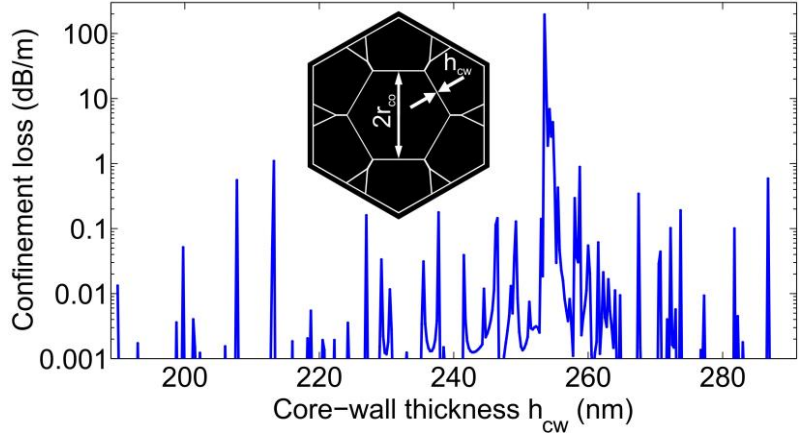

Fig. 2: The loss of $LP_{01}$-like modes in the simplified kagomé-structure shown in the inset, plotted against core-wall thickness. The results were calculated numerically using FEM.

## 3. Experimental results

A fiber laser emitting at 1121 nm was frequency-quadrupled to produce up to 20 mW at 280 nm [2]. After spatial filtering to clean up the beam profile, the light was focused to a beam-waist of 15 μm using a lens of focal length 75 mm and launched into the fiber.

*3.1 Loss, bending and polarization properties*
The loss was measured using a multiple cut-back technique, without changing the fiber in-coupling and keeping the fiber as straight as possible so as to minimize bending loss. The results are shown in Fig. 3(a). Fiber A had a measured core-wall thickness in the range ~220-260 nm (Fig. 1(b)), and yielded a loss of ~3 dB/m, which is in agreement with the simulated value for a core-wall thickness of ~250-260 nm (Fig. 2). For fiber B the measured loss was ~0.8 dB/m, in agreement with the simulated loss (~1 dB/m) for a core-wall thickness in the range 190-220 nm.

Several mW of power at 280 nm was transmitted, representing typically ~48% of the launched power for fiber A (~1 m fiber length) and ~70% for fiber B (~1.3 m length). Bending was found to cause large variations in the transmitted power, probably due to enhanced leakage into the cladding and coupling to high-loss higher-order modes. If the fiber is either kept straight or the bending is not changed the output remained stable and, depending on the fiber alignment, the transmission can reach its maximum value for bend-radii greater than ~20 cm. To quantify the bend-sensitivity the output power was measured while winding fiber A around a mandrel. The transmitted power exhibited fluctuations depending on the exact position and twist of the fiber. The output power was therefore recorded while rearranging the fiber and the maximum measured power was used to derive the bending loss. For 5 turns at a bending radius of 16 mm the loss was smaller than 2.9 dB/turn, which is equivalent to more than ~25% transmission through 1 m of fiber including one full turn with 16 mm bending radius. For 40 mm bending radius and up to 3 turns the measured loss was smaller than 2.7 dB/turn.

It was also found that the transmission depends on the polarization state of the light. This is normally not expected in kagomé-PCFs with perfect six-fold symmetry, but as was shown with simulations in the previous section (see Fig. 2), the loss is sensitive to nm-scale variations in core-wall thickness. Azimuthal variations in core-wall thickness can therefore result in polarization-dependent loss. In Fig. 3(b) the normalized transmission along 1 m lengths of fiber is plotted while varying the linear input polarization. There is a maximum ~20% (~1 dB/m) difference in output power between orthogonal polarizations for fiber A, and a maximum ~10% (~0.5 dB/m) for fiber B. To check that this variation was really due to the structure of the fiber, the measurement was repeated after turning the input end of the fiber by ~90°. As seen in Fig. 3(b) this simply rotated the plot by 90°, supporting that the effect was indeed due to the structure of the fiber.

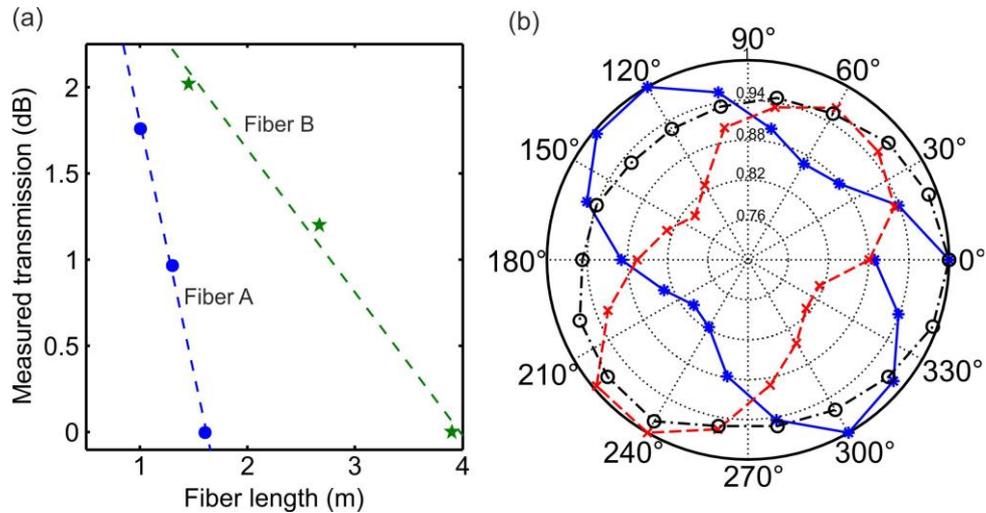

Fig. 3. (a) Cut-back loss measurement for fiber A (blue dots) and fiber B (green stars) at a wavelength of 280 nm. The linear fits correspond to 2.9 dB/m (blue, fiber A) and 0.8 dB/m (green, fiber B). (b) Normalized transmission (center = 0.7) plotted radially versus orientation of the linearly polarized input light for fibers 1 m in length. After making the first measurement on fiber A (solid, blue) the input end of the fiber was turned by ~90° and the measurement repeated (red, dashed). The elliptical shape of the polarization-dependent transmission follows the fiber orientation and exhibits a consistent maximum polarization-dependent loss of ~20%. For the blue plot there was a slight misalignment during the measurement causing an error in the registration at 0° and 360°. An equivalent measurement for fiber B (black, dash-dotted) showed a maximum polarization dependent loss of ~10% .

*3.2 Output beam quality*

The robustness of the single-mode guidance was tested by monitoring the output beam profile while translating the input coupling beam across the input face of the fiber. Little evidence of higher-order modes was observed (Fig. 4). We note that in Ref. [14] the 355 nm light output from a fiber with ~30 μm core diameter showed a highly multimoded pattern. The kagomé-PCFs considered here are effectively single-mode due to a smaller core diameter (~19 μm) and therefore higher loss for higher-order modes, which means that they act as simple mode-cleaners.

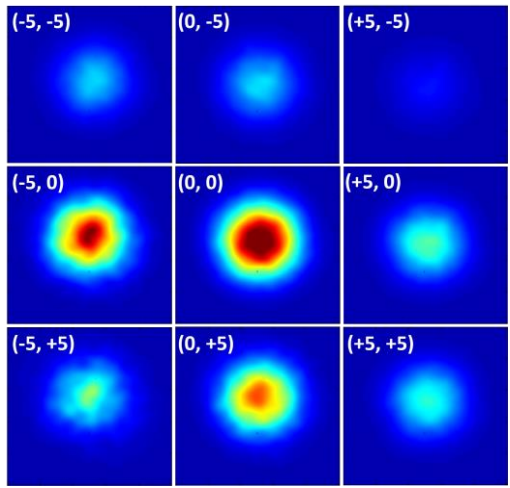

Fig. 4. Measured near-field intensity profiles from fiber A for different transverse positions of the input beam. The coordinates refer to the horizontal and vertical displacements (in μm) of the focused laser spot from the core center.

*3.3 Lifetime investigation*

As mentioned in the introduction there is a current need for optical fibers that can deliver UV light with high beam quality without degradation due to UV-induced damage to the glass core. Experiments with a solid-core PCF showed that the transmission of 280 nm light dropped by more than 80% after 3 hours at a CW power level of only 2 mW, in line with previous reports [11].

FEM calculations indicate that less than 0.01% of the light in a perfect kagomé-PCF is guided in the glass [13], which suggests that UV-induced damage should be largely eliminated. To investigate this we launched 15 mW of 280 nm CW light into fiber A and monitored the transmission continuously over time. Within measurement error we could detect no change in the transmission after 100 hours.

## 4. Applications

In order to study the applicability of the fiber in trapped ion experiments, an intensity stabilization set-up was implemented. The fiber ends were fixed with sticky tape in the V-groove of an aluminum block and the fiber was used to replace a periscope system that connected two stacked platforms. The distance from the output face of the fiber to the trapped ion was kept as short as possible so as to minimize residual pointing fluctuations from air currents.

Pointing instabilities lead to intensity fluctuations at the position of the ion and can therefore limit the laser control of its internal state. For coherent manipulation of the ion we used a Raman beam configuration to excite the $|F = 3, m_F = 3\rangle = |\downarrow\rangle \leftrightarrow |F = 2, m_F = 2\rangle = |\uparrow\rangle$ transition in the $^2S_{1/2}$ state of a $^{25}Mg^+$ ion via near-resonant Raman coupling detuned by 9.2 GHz to the $^2P_{3/2}$ excited state [2]. Here, F denotes the total angular momentum and $m_F$ its projection along the magnetic field direction. These so-called qubit states are separated by a hyperfine splitting of 1.789 GHz and further split by an external bias magnetic field that lifts the degeneracy of the magnetic sub-states. The Raman beams were generated by frequency quadrupling the output of a fiber laser at 1118 nm to 280 nm. After splitting the UV light into two beams each one passed through several acousto-optical modulators (AOMs) to bridge the frequency difference between the two qubit states. The complete optical set-up and the level scheme of $^{25}Mg^+$ are described in [2]. Since no single-mode fibers were previously available at a wavelength of 280 nm, the original set-up consisted of free-space beam-paths approximately 5 m long. Air turbulence and small vibrations of mirrors and other optical components led to noticeable pointing fluctuations at the position of the ion. Furthermore, the AOMs distorted the beam shape, which caused additional intensity gradients across the beam profile. These were cleaned to a near-Gaussian transverse mode shape using the hollow core kagomé-PCFs discussed here.

The effect of beam-pointing fluctuations was investigated by measuring the signal contrast for laser-driven coherent internal state oscillations both with and without a kagomé-PCF. To this end we recorded the internal state of the ion after coupling the two hyperfine states via the Raman lasers for different pulse durations (Rabi flopping). The excitation probability is determined by averaging the result of approximately 250 repetitions of the experiment per Raman interaction time. If the position of the laser beam and therefore the light intensity on the ion fluctuates, the Rabi frequency (which is a function of the laser intensity) will change between subsequent experiments. During the averaging process this results in a reduction of the measured Rabi oscillation contrast, which can be described as a damped oscillation of the ground state population [20]:

$$P_{|\downarrow\rangle}(t) = \tfrac{1}{2}\left[1 + e^{-\gamma t}\cos(\Omega_0 t)\right] \quad (2)$$

where $\Omega_0$ is the mean Rabi frequency, $\gamma = \Gamma\Omega_0^2/2$ the decay rate of the Rabi oscillations, and $\Gamma$ a scale-factor for the intensity noise. Competing effects that reduce the Rabi oscillation contrast further are the motional excitation of the ion in the trap due to its non-zero temperature, and off-resonant scattering of the Raman beams [21]. Off-resonant scattering is not a fundamental limitation, since it can be mitigated by detuning the Raman resonance further from the atomic transition. Preparation of the ion close to the motional ground state of the $|\downarrow\rangle$-state, via fast Raman sideband cooling [22] at the beginning of each experimental cycle, eliminates the influence of motional excitation on the Rabi-flopping contrast. After cooling to a mean motional excitation of $\bar{n} = 0.02 \pm 0.02$, where $\bar{n}$ is the mean motional quantum number, we applied the Raman coupling for different pulse durations while actively stabilizing the intensity of the Raman lasers using a sample-and-hold intensity stabilization circuit.

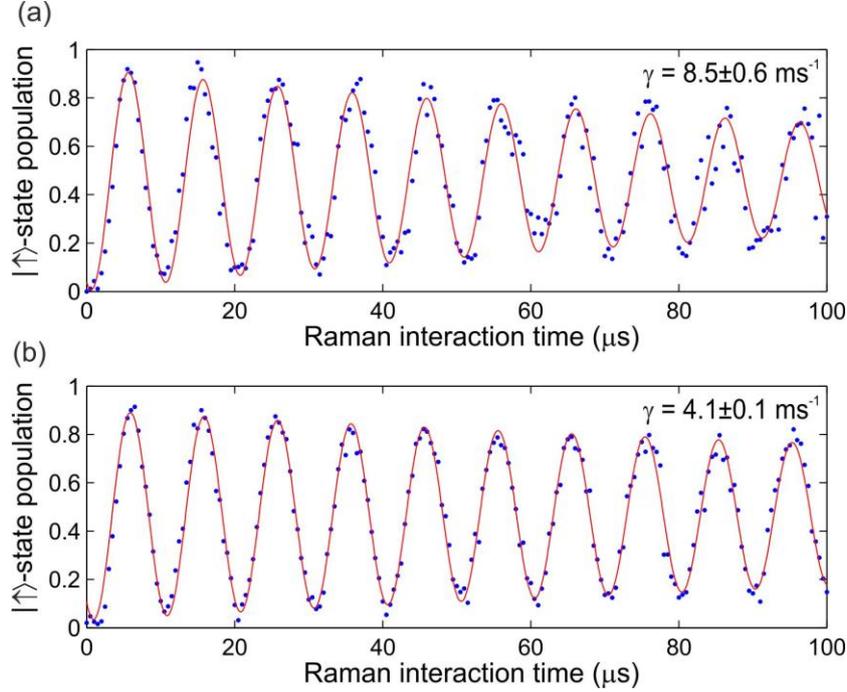

Fig. 5. Raman Rabi oscillations (a) without kagomé-PCF and (b) with kagomé-PCF in the set-up. The decay rate is extracted from a fit to Eq. (2) and the mean value over 7 measurements is displayed in the corresponding graphs. The residual decay is dominated by off-resonant excitation from the Raman lasers due to the limited Raman detuning of 9.2 GHz.

Figure 5 shows experimental data for the Rabi flopping curves with and without the kagomé-PCF, together with a corresponding fit according to Eq. (2). The extracted decay rate γ is shown in Fig. 5 and is averaged over 7 different measurements in each configuration. The resulting weighted average of the decay rates is 4.1±0.1 ms$^{-1}$ with, and 8.5±0.6 ms$^{-1}$ without, the kagomé-PCF in place. The relatively strong residual decay is dominated by off-resonant excitation of the ion to the electronically excited $^2P_{3/2}$ state, due to the limited Raman detuning of 9.2 GHz. This is confirmed in an independent experiment in which we initially prepared the ion in one of the two qubit states and then applied the Raman lasers detuned to the two-photon resonance by half the distance to the next resonance (1.1 MHz). Subsequent measurements of the internal state of the ion yielded the decay rates of the two states due to off-resonant excitation $\gamma_{|\downarrow\rangle} = 1.1 \pm 0.1$ ms$^{-1}$ and $\gamma_{|\uparrow\rangle} = 4.4 \pm 0.4$ ms$^{-1}$, which are comparable to the residual decay rate of the Rabi oscillation with the kagomé-PCF in the set-up in Fig. 5(b). Further investigations at a larger Raman detuning would be necessary to determine the ultimate limit of residual Rabi frequency fluctuations.

An additional advantage of adding the kagomé-PCF to this type of set-up for single ion experiments is that any realignment of the beam-path before the fiber does not shift the focal position on the ion, which is typically cumbersome to achieve in free-space set-ups.

## 5. Conclusions

Kagomé-style photonic crystal fibers provide high-quality single-mode transmission at 280 nm wavelength with losses of ~1 dB/m. The transmitted beam is free of laser pointing instabilities and unlike in solid-core fibers there is no perceptible drop in transmission due to UV-induced damage, even after 100 hours of operation at 15 mW. We note that an alternative approach, based on hydrogen-loading of a solid-core photonic crystal fiber, has recently been reported [23].

**Acknowledgements**

F. Gebert, Y. Wan and P.O. Schmidt acknowledge support by DFG through QUEST and SCHM2678/3-1. Y. Wan acknowledges support from IGSM.